\documentclass[12pt,letterpaper]{article}
\pdfoutput=1
\usepackage{amssymb}
\usepackage{amsmath}
\usepackage{fourier-orns}
\usepackage{amscd}
\usepackage{graphicx}
\usepackage{hyperref}
\def\be{\begin{equation}}
\def\ee{\end{equation}}
\def\frc#1#2{\relax\ifmmode{\textstyle\frac{#1}{#2}} 
                    \else$\frac{#1}{#2}$\fi}         

\def\rI{{{}_{\rm I}}}
\def\rJ{{{}_{\rm J}}}

\def\hk{{\hat k}}
\def\hi{{\hat\imath}}

\def\fracm#1#2{\hbox{\large{${\frac{{#1}}{{#2}}}$}}}

\def\pp{{\mathchoice
          {
              \kern 1pt%
              \raise 1pt
              \vbox{\hrule width5pt height0.4pt depth0pt
                    \kern -2pt
                    \hbox{\kern 2.3pt
                          \vrule width0.4pt height6pt depth0pt
                          }
                    \kern -2pt
                    \hrule width5pt height0.4pt depth0pt}%
                    \kern 1pt
           }
            {
              \kern 1pt%
              \raise 1pt
              \vbox{\hrule width4.3pt height0.4pt depth0pt
                    \kern -1.8pt
                    \hbox{\kern 1.95pt
                          \vrule width0.4pt height5.4pt depth0pt
                          }
                    \kern -1.8pt
                    \hrule width4.3pt height0.4pt depth0pt}%
                    \kern 1pt
            }
            {
              \kern 0.5pt%
              \raise 1pt
              \vbox{\hrule width4.0pt height0.3pt depth0pt
                    \kern -1.9pt  
                    \hbox{\kern 1.85pt
                          \vrule width0.3pt height5.7pt depth0pt
                          }
                    \kern -1.9pt
                    \hrule width4.0pt height0.3pt depth0pt}%
                    \kern 0.5pt
            }
            {
              \kern 0.5pt%
              \raise 1pt
              \vbox{\hrule width3.6pt height0.3pt depth0pt
                    \kern -1.5pt
                    \hbox{\kern 1.65pt
                          \vrule width0.3pt height4.5pt depth0pt
                          }
                    \kern -1.5pt
                    \hrule width3.6pt height0.3pt depth0pt}%
                    \kern 0.5pt
            }
        }}

\def\mm{{\mathchoice
                       {
                             \kern 1pt
               \raise 1pt    \vbox{\hrule width5pt height0.4pt depth0pt
                                  \kern 2pt
                                  \hrule width5pt height0.4pt depth0pt}
                             \kern 1pt}
                       {
                            \kern 1pt
               \raise 1pt \vbox{\hrule width4.3pt height0.4pt depth0pt
                                  \kern 1.8pt
                                  \hrule width4.3pt height0.4pt depth0pt}
                             \kern 1pt}
                       {
                            \kern 0.5pt
               \raise 1pt
                            \vbox{\hrule width4.0pt height0.3pt depth0pt
                                  \kern 1.9pt
                                  \hrule width4.0pt height0.3pt depth0pt}
                            \kern 1pt}
                       {
                           \kern 0.5pt
             \raise 1pt  \vbox{\hrule width3.6pt height0.3pt depth0pt
                                  \kern 1.5pt
                                  \hrule width3.6pt height0.3pt depth0pt}
                           \kern 0.5pt}
                       }}

\def\ad{{\kern0.5pt
                   \alpha \kern-5.05pt \raise5.8pt\hbox{$\textstyle.$}\kern
0.5pt}}

\def\bd{{\kern0.5pt
                   \beta \kern-5.05pt \raise5.8pt\hbox{$\textstyle.$}\kern
0.5pt}}

\def\qd{{\kern0.5pt
                   q \kern-5.05pt \raise5.8pt\hbox{$\textstyle.$}\kern
0.5pt}}
\def\Dot#1{{\kern0.5pt
     {#1} \kern-5.05pt \raise5.8pt\hbox{$\textstyle.$}\kern
0.5pt}}

\catcode`@=11
\def\un#1{\relax\ifmmode\@@underline#1\else
        $\@@underline{\hbox{#1}}$\relax\fi}
\catcode`@=12

\def\a{\alpha}
\def\b{\beta}

\def\e{\epsilon}

\def\g{\gamma}

\def\l{\lambda}
\def\m{\mu}
\def\n{\nu}

\def\p{\pi}

\def\t{\tau}

\def\D{\Delta}

\def\L{\Lambda}

\def\S{\Sigma}

\def\ce{{\cal E}}

\def\cn{{\cal N}}

\def\ct{{\cal T}}

\def\cx{{\cal X}}
\def\cy{{\cal Y}}
\def\cz{{\cal Z}}
 
\def\dslash{\not{\hbox{\kern-2pt $\partial$}}}
\def\Dslash{\not{\hbox{\kern-4pt $D$}}}
\def\pslash{\not{\hbox{\kern-2.3pt $p$}}}
 \newtoks\slashfraction
 \slashfraction={.13}
 \def\slash#1{\setbox0\hbox{$ #1 $}
 \setbox0\hbox to \the\slashfraction\wd0{\hss \box0}/\box0 }
 
\font\ro=cmsy10                          
\def\kcr{{\hbox{\ro \char'170}}}                
\def\ktl{{\hbox{\ro \char'170}}}        
\def\ktr{{\hbox{\ro \char'170}}}        
\def\kbl{{\hbox{\ro \char'170}}}        
\def\kbr{{\hbox{\ro \char'170}}}        

\def\plpl{\raise-2pt\hbox{$\raise3pt\hbox{$_+$}\hskip-6.67pt\raise0.0pt
\hbox{$^+$}\hskip 0.01pt$}}
\def\mimi{\raise-2pt\hbox{$\raise3pt\hbox{$_-$}\hskip-6.67pt\raise0.0pt
\hbox{$^-$}\hskip 0.01pt$}} 

\def\bo{{\raise.15ex\hbox{\large$\Box$}}}               
\def\pa{\partial}                                       
\def\TH{{\raise.2ex\hbox{$\displaystyle \bigodot$}\mskip-4.7mu \llap H \;}}
\def\face{{\raise.2ex\hbox{$\displaystyle \bigodot$}\mskip-2.2mu \llap {$\ddot
        \smile$}}}                                      

\def\Tilde#1{\widetilde{#1}}                    
\def\Hat#1{\widehat{#1}}                        
\def\leftrightarrowfill{$\mathsurround=0pt \mathord\leftarrow \mkern-6mu
        \cleaders\hbox{$\mkern-2mu \mathord- \mkern-2mu$}\hfill
        \mkern-6mu \mathord\rightarrow$}
\def\dvec#1{\vbox{\ialign{##\crcr
        \leftrightarrowfill\crcr\noalign{\kern-1pt\nointerlineskip}
        $\hfil\displaystyle{#1}\hfil$\crcr}}}           
\def\dt#1{{\buildrel {\hbox{\LARGE .}} \over {#1}}}     

\def\fracm#1#2{\hbox{\large{${\frac{{#1}}{{#2}}}$}}}
\def\frac#1#2{{\textstyle{#1\over\vphantom2\smash{\raise.20ex
        \hbox{$\scriptstyle{#2}$}}}}}                   
\def\sfrac#1#2{{\vphantom1\smash{\lower.5ex\hbox{\small$#1$}}\over
        \vphantom1\smash{\raise.4ex\hbox{\small$#2$}}}} 
\def\bfrac#1#2{{\vphantom1\smash{\lower.5ex\hbox{$#1$}}\over
        \vphantom1\smash{\raise.3ex\hbox{$#2$}}}}       
\def\afrac#1#2{{\vphantom1\smash{\lower.5ex\hbox{$#1$}}\over#2}}    
\def\on#1#2{\mathop{\null#2}\limits^{#1}}               

\def\pa{\partial}      
\newcommand{\bm}[1]{\mbox{\boldmath$#1$}}

\def\ad{{\dot{\alpha}}}
\def\bd{{\dot{\beta}}}

\font\ro=cmsy10                          
\def\kcr{{\hbox{\ro \char'170}}}                
\def\ktl{{\hbox{\ro \char'170}}}        
\def\ktr{{\hbox{\ro \char'170}}}        
\def\kbl{{\hbox{\ro \char'170}}}        
\def\kbr{{\hbox{\ro \char'170}}}        

\parskip=\medskipamount                 
\thicklines                         

\def\border{                                            
        \setlength{\unitlength}{1mm}
        \newcount\xco
        \newcount\yco
        \xco=-21
        \yco=12
        \begin{picture}(140,0)
        \put(\xco,\yco){$\ktl$}
        \advance\yco by-1
        {\loop
        \put(\xco,\yco){$\kcr$}
        \advance\yco by-2
        \ifnum\yco>-240
        \repeat
        \put(\xco,\yco){$\kbl$}}
        \xco=158
        \yco=12
        \put(\xco,\yco){$\ktr$}
        \advance\yco by-1
        {\loop
        \put(\xco,\yco){$\kcr$}
        \advance\yco by-2
        \ifnum\yco>-240
        \repeat
        \put(\xco,\yco){$\kbr$}}
        \put(-20,13){\tiny **University of Maryland * Center for String and
         Particle  Theory* Physics Department***University of Maryland *Center
        for String and Particle  Theory** }
        \put(-20,-241.5){\tiny **University of Maryland * Center for String and
         Particle  Theory* Physics Department***University of Maryland *Center
        for String and Particle  Theory** }
        \end{picture}
        \par\vskip-8mm}

\def\headpic{                                           
        \indent
        \setlength{\unitlength}{.4mm}
        \thinlines
        \par
        \begin{picture}(29,16)
        \put(165,16){\line(1,0){4}}
        \put(170,16){\line(1,0){4}}
        \put(180,16){\line(1,0){4}}
        \put(175,0){\line(1,0){4}}
        \put(180,0){\line(1,0){4}}
        \put(185,0){\line(1,0){4}}
        \put(169,0){\line(0,1){16}}
        \put(170,0){\line(0,1){16}}
        \put(179,0){\line(0,1){16}}
        \put(180,0){\line(0,1){16}}
        \put(184,0){\line(0,1){16}}
        \put(185,0){\line(0,1){16}}
        \put(169,16){\oval(8,32)[bl]}
        \put(170,16){\oval(8,32)[br]}
        \put(179,0){\oval(8,32)[tl]}
        \put(185,0){\oval(8,32)[tr]}
        \end{picture}
        \par\vskip-6.5mm
        \thicklines}
\def\endtitle{\end{quotation}\newpage}                  

\newskip\humongous \humongous=0pt plus 1000pt minus 1000pt
\def\caja{\mathsurround=0pt}
\def\eqalign#1{\,\vcenter{\openup2\jot \caja
        \ialign{\strut \hfil$\displaystyle{##}$&$
        \displaystyle{{}##}$\hfil\crcr#1\crcr}}\,}
\newif\ifdtup

\renewcommand{\=}{~=~}

\renewcommand{\L}{{\rm L}}
\newcommand{\R}{{\rm R}}
\renewcommand{\D}{{\rm D}}

\renewcommand{\bar}[1]{\overline{#1}}

\begin{document}
\numberwithin{equation}{section} 
\setcounter{page}{0}
\thispagestyle{empty}

\def\dt#1{\on{\hbox{\bf .}}{#1}}                
\def\Dot#1{\dt{#1}}

\def\gfrac#1#2{\frac {\scriptstyle{#1}}
        {\mbox{\raisebox{-.6ex}{$\scriptstyle{#2}$}}}}
\def\gg{{\hbox{\sc g}}}
\border\headpic {\hbox to\hsize{\today \hfill
{UMDEPP-013-019}}}
\par \noindent
{ \hfill{arXiv:1312.2000 [hep-th]}}
\par

\setlength{\oddsidemargin}{0.3in}
\setlength{\evensidemargin}{-0.3in}
\begin{center}
\vglue .08in
{\large\bf  Reduction Redux of Adinkras }\\[.5in]

S.\, James Gates, Jr.\footnote{gatess@wam.umd.edu}${}^{\dagger}$,
Stephen Randall\footnote{stephenlrandall@gmail.com}$^{\dagger}$,
and Kory Stiffler\footnote{kmstiffl@iun.edu}${}^{*}$
\\[0.3in]
${}^\dag${\it Center for String and Particle Theory\\
Department of Physics, University of Maryland\\
College Park, MD 20742-4111 USA}\\[0.1in]
and
\\[0.1in]
${}^*${\it Department of Chemistry, Physics, \& Astronomy\\
College of Arts and Sciences\\
Indiana University Northwest\\
Gary, Indiana 46408 USA}
\\[0.9in] $~$
{\bf ABSTRACT}\\[.01in]
\end{center}
\begin{quotation}
{We show performing general ``0-brane reduction'' along an arbitrary
fixed direction in spacetime and applied  to the starting point of minimal, 
off-shell 4D, $\cal N$ $=$ 1 irreducible supermultiplets, yields adinkras 
whose adjacency matrices are among some of the special cases 
proposed by Kuznetsova, Rojas, and Toppan.  However, these more
general reductions also can lead to `Garden Algebra' structures beyond 
those described in their work.  It is also shown that for light-like directions,
reduction to the 0-brane breaks the equality in the number 
of fermions and bosons for dynamical theories.  This implies that 
light-like reductions 
should be done instead to the space of 1-branes or equivalently to the worldsheet.}

\endtitle

\setlength{\oddsidemargin}{0.3in}
\setlength{\evensidemargin}{-0.3in}

\setcounter{equation}{0}
\section{Introduction}

$~~~$ A number of years ago \cite{GRana}, we began the study
of a ubiquitous mathematical sub-structure that appears in all off-shell 
linear representations of supersymmetry in one and two dimensions.   
The topic was developed in a series of papers which uncovered 
increasingly complex intricacies such as a series of mapping 
operations acting on these structures.  In the works of \cite{ENUF},
a procedure,``0-brane projection'', was described whereby these 
same structures could be derived from supersymmetric theories
in all higher dimensions.  Eventually, these structures were named 
``adinkras'' \cite{adinkra1}.

It is rather simple to prove that when any standard field-theoretic 
formulation of a supersymmetric field theory is examined under 
such a projection, forcing all dynamical fields to {\em {only}} depend 
on a single temporal coordinate, this results in the revealing of adinkras. 
So these networks appear to be {\em {universal}} to all supersymmetric 
field theories.  Shortly after the christening, there was assembled an
interdisciplinary collaboration of mathematicians and physicists (the 
`DFGHILM group'\footnote{A large number of these publications can 
be identified by using the webpage resource at \newline $~~~~~\,$  
\url{http://math.bard.edu/DFGHIL/index.php?n=Main.Publications} on-line.}) 
which worked diligently to define more rigorously the properties of 
adinkras.  

By now there has been established a substantial literature that treats 
aspects of this approach to understanding the fundamental structure of 
off-shell spacetime supersymmetric representations.  Along the way 
many unexpected developments have occurred.  The most recent such 
surprise \cite{geoadnk} is the discovery that adinkras are directly related 
to super Riemann surfaces with {\em {no}} reference to Superstring/M-Theory!  
Adinkras are equivalent to very special super Riemann surfaces with 
divisors.

We have long advocated \cite{ENUF} that adinkras may be regarded 
as the equivalent of genes for biological systems.  That is, we have 
suggested it is possible to begin {\em {solely}} with adinkras, which 
are intrinsically one dimensional networks, and then reconstruct higher 
dimensional supersymmetric representations from this starting point.  
We have named this process ``SUSY holography'' as should it succeed, 
this would imply that the information to reconstruct the higher dimensional 
supersymmetric systems must in some way be {\em {holographically 
encoded}} within the starting point of one dimensional adinkra networks.  

Within the last year, two presentations have advanced the possibility 
of creating a proof of the ``SUSY holography conjecture'' at least regarding 
the relation between 1D, $N$ = 4 adinkras and 4D, $\cal N$ 
= 1 classes of theories.  

First, it was established \cite{KIAS} that given some specific 
1D, $N$ = 4 adinkras, their associated adjacency matrices necessarily carry 
a representation of an SU(2)\,$\otimes$\,SU(2) algebra\footnote{The presence
of this SU(2)\,$\otimes$\,SU(2) algebra was first observed in the work of 
\cite{G-1} where it was \\ $~~~\,~~$ noted that while off-shell representations
are representation of this algebra, on-shell represen- \\ $~~~\,~~$ tations 
only carry representations of an SU(2)\,$\otimes$\,SO(2) algebra.}.  There is 
a group theoretical understanding of this result that leads to the conclusion 
that for a general adinkra with $N$ colors, the associated adjacency matrices 
will carry some representation of SO($N$).  Since locally the algebra of SO(4) 
is isomorphic to an SU(2)\,$\otimes$\,SU(2) algebra, the demonstrated results 
consequently follow. 

Though it appears not generally appreciated, the covering algebra of the SO(1,\,3)
Clifford algebra also carries a representation of an SU(2)\,$\otimes$\,SU(2) algebra.  
It is possible to identify these two algebras as one and the same structure and 
re-construct, solely from the 1D, $N$ = 4 adinkras, matrices in 
the covering algebra of the SO(1,\,3) $\g$-matrices.  Thus, information about the 
SO(1,\,3) spin bundle associated with a  4D, $\cn$ = 1 supermultiplet can indeed 
be encoded within the network of adinkras!

Second, it was established \cite{permutadnk} that the sets of all possible quartets 
of 4 $\times$ 4 matrices that can be used to describe the `L-matrices' and related 
`R-matrices' associated with minimal off-shell 4D, $\cn$ = 1 supermultiplets (the 
chiral, vector, and tensor supermultiplets) can be identified with the 384 elements 
of the Coexter Group $BC_4$.  Taking absolute values of the elements of these 
matrices then leads to a further identification with $S_4$, the permutation group 
of four elements.  

Next, it was shown there exists a discrete transformation (acting on these adinkras 
identified with the elements of the Coexter Group $BC_4$) with the property of 
defining a class structure of three distinct parts.  It was then argued 
that these three sub-classes should be identified with the respective three distinct 
minimal off-shell representations of 4D, $\cn$ = 1 supersymmetry and that the 
operation provides (in the space of 1D, $N$ = 4 adinkras) a realization 
of a `one dimensional Hodge star operation' acting on forms in the covering algebra 
of the SO(1,\,3) Clifford algebra!

These two results together establish a fairly well defined path for the reconstruction
of 4D, $\cn$ = 1 supermultiplets from one dimensional four-color adinkras:  \newline 
$~~~~$
(a.) using the putative and proposed `one dimensional  Hodge star   \newline 
       $~~~~\,~~~~~~$ operation' any such adinkra can be associated with one of
       \newline $~~~~\,~~~~~~$  the minimal 4D, $\cn$ = 1 
       supersymmetric representations, and \newline $~~~~$
(b.) the SU(2)\,$\otimes$\,SU(2) content associated with the given adinkra de-
       \newline $~~~~\,~~~~~~$ termines the SO(1,\,3) spin bundle representation 
       carried by the  \newline $~~~~\,~~~~~~$ nodes of the adinkra. \newline 
       \noindent
Of course, more work understanding the details of these steps is required in the
general case as well as a complete understanding of the filters that allow completely
consistent dimensional enhancement.  But the results in hand provide the essence 
of an existence proof that we believe can  be successfully constructed.       
 
All these results are within the context of adinkras associated with our traditional ``0-brane 
reduction'' of 4D, $\cn$ = 1 supermultiplets.  However, in this work, we turn to a different 
question.  Up to this point previously, whenever we have constructed adinkras from higher 
dimensional supersymmetric field theories, the single bosonic coordinate in the adinkra 
was associated with a purely time-like direction.  It is a very natural question to ask, ``What might 
occur if instead the reduction was done with respect to an arbitrary constant 4-vector direction
among the spacetime coordinates?''  This is the question probed in this current work.

We shall show in the following work, using the familiar off-shell minimal 4D, $\cn$ = 1 
SUSY representations, that
``0-brane reduction'' utilizing a non-purely time-like direction yields adinkras whose associate
L-matrices and associated R-matrices form representations that are generalizations of the
``Garden Algebras'' introduced in work of \cite{GRana}.  The authors Kuznetsova, Rojas, 
and Toppan have previously introduced generalizations  \cite{EtAlToppan} of the
``Garden Algebras'' some time ago.  We shall indeed see that ``0-brane reduction'' 
utilizing a space-like direction can yield structures first suggested by these authors.
However, we shall also see that the ``0-brane reduction'' utilizing a general 
space-like 
direction produces generalizations even beyond those considered in their work.


$~~~~$ \section{General 0-Brane Reduction}
\label{section2}

$~~~~$ The starting point to revealing the adinkra sub-structure of higher dimensional
supersymmetric theories in Minkowski space begins with the usual spacetime four-gradient 
operator $\pa_{\m}$
\be  \eqalign{
\pa_{\m} \,&=\, \left(\,  {{\pa} \over {\partial t}} \, , \, \,   {{\pa} \over {\partial x}} \,
,\, \,  {{\pa} \over {\partial y}} \, ,\,\,   {{\pa} \over {\partial z}} \,  \right)   
 ~=~ {\cal T}_{\mu} \, {{\pa} \over {\partial t}} \, +\, {\cal X}_{\mu}  {{\pa} \over {\partial x}}
\, + \, {\cal Y}_{\mu} {{\pa} \over {\partial y}} \, + \,  {\cal Z}_{\mu} {{\pa} \over {\partial z}} \, 
~~~,
}   \label{DerV}
\ee
where the constant four-vectors ${\cal T}_{\mu}$,  ${\cal X}_{\mu}$,
${\cal Y}_{\m}$, and  ${\cal Z}_{\m}$ respectively denote
\be \eqalign{
{\cal T}_{\mu}  \,&=~ (\, 1, \,0,\, 0 ,\, 0 \, ) ~~~,~~~
{\cal X}_{\mu} \,=~ (\, 0, \,1,\, 0 ,\, 0 \, ) ~~~, \cr
{\cal Y}_{\m} \,&=~ (\, 0, \,0,\, 1 ,\, 0 \, ) ~~~,~~~
{\cal Z}_{\m} \,=~ (\, 0, \,0,\, 0 ,\, 1 \, ) ~~~.
}    \label{VectS}
\ee
Next a single real parameter $\t$ may be introduced via the equations below
\be  \eqalign{
&{{\pa} \over {\partial t}}   ~=~ \cos \a \, {{\pa} \over {\partial \t}}  ~~~~,  ~~~~~~~~~~~~
~~\,~~~
{{\pa} \over {\partial x}} ~=~    \sin \a \,  \sin \b  \,\cos \g  \, {{\pa} \over {\partial \t}}  ~~~~,  \cr
&{{\pa} \over {\partial y}}   ~=~   \sin \a \, \sin \b  \,  \sin \g  \, {{\pa} \over {\partial \t}}  ~~~~,  ~~~~
{{\pa} \over {\partial z}} ~=~    \sin \a \, \cos \b  \, {{\pa} \over {\partial \t}}  ~~~~,  \cr
}    \label{DerVs}
\ee
whereupon the  four-gradient operator $\pa_{\m}$ takes the restricted form $\pa_{\m}$
$=$ $\ell_{\mu}\, \pa_{\t}$ or more explicitly
\be  \eqalign{
\pa_{\m} &= \left[ \cos\a   \, {\cal T}_{\mu}   +  \sin\a   \sin\b  \cos\g   \, {\cal X}_{\mu} 
 +     \sin\a  \sin\b    \sin \g  \, {\cal Y}_{\mu}   +    \sin \a  \cos \b   
\, {\cal Z}_{\mu}  \right] {{\pa} \over {\partial \t}} ~.  }   
  \label{DerV-t}
\ee
This restricted form of the spacetime four-gradient operator clearly generates motion 
in the Minkowski space along a straight line described by the four-vector $\ell_{\m}$.
We have named this process `0-brane reduction.'  It has long been our assertion that
the group theoretic structures obtained from 0-brane reduction in the context of 
spacetime supersymmetric representation theory plays the same role as the Wigner 
`little group'  for non-supersymmetric representation theory.   

 In a four-dimensional $\cal N$ = 1 theory, the anti-commutator algebra
 for the supersymmetry generator can be written as
 \be  {
\{ \, {\rm Q}_a  \,,\,  {\rm Q}_b  \,  \} \,  
~=~  i\, 2 \,  \, (\gamma^\mu){}_{a \,b}\,  \partial_\mu  ~~~,
}  \label{SUSYalg}
\ee
which under 0-brane reduction becomes
 \be  {
\{ \, {\rm Q}_a  \,,\,  {\rm Q}_b  \,  \} \,  
~=~  i\, 2 \,  \, (\gamma^\mu){}_{a \,b}\,  \ell_{\mu}\, \pa_{\t}  
~=~  i\, 2 \,  \, (\gamma \cdot \ell){}_{a \,b} \, \pa_{\t} 
~~~.
 }  \label{SUSYalgT}
\ee

At this stage, it does not appear that any particular 0-brane projection (for arbitrary 
values of the parameters $\a$, $\b$, and $\g$) possesses any distinguished behavior 
from any other projection.  However, in the following we will see that though the 
change above is mild, when the questions of dynamics are engaged, depending 
on whether $\ell$ is time-like or space-like versus light-like, subtle differences do 
emerge.

As discussed in \cite{G-1}, our gamma matrices are chosen to be 
real and explicitly given by
 \be   \eqalign{
&~~~~~~~~~~~~~~~~~~~ {(\gamma^0)}  = i ( \sigma^3
 \otimes \sigma^2  ) 
~~~~,~~~~~~ {(\gamma^1)}  = ({\bf I}_2 
\otimes \sigma^1 ) ~~~~~, \cr
&~~~~~~~~~~~~~~~~~~~ {(\gamma^2)}  = (\sigma^2 
\otimes \sigma^2 ) ~~~~~,~~~~~~ 
  {(\gamma^3)}  = ({\bf I}_2 
\otimes \sigma^3  )  ~~~~~,
}   \label{G-matrX}
\ee
thus describing a mostly plus Minkowski spacetime metric $\eta{}^{\mu \, \nu} $
that appears in
\be {
\gamma^\mu \, \gamma^\nu ~+~  \gamma^\nu \, \gamma^\mu ~=~
2 \, \eta{}^{\mu \, \nu} \, {\bf I}_4 
}  \label{G-matrX2}
\ee
so that the purely imaginary `gamma-five' matrix takes the form
\be {~~~~~~~
 {(\gamma^5)}  ~=~ i \, \g^0 \, \g^1 \, \g^2 \, \g^3
  ~=~  -(\sigma^1 \otimes \sigma^2 ) ~~~~~. 
 } \label{G5-matrX3}
\ee
From the definitions above, it follows that the generators of spatial rotations
$\S^{i \, j} \, =\, i/4 [ \,  \gamma^i \, ~,~ \gamma^j \, ]$
take the explicit forms
\be {
\S^{1 \,2} ~=~  \fracm 12 \,  (\sigma^2 
\otimes \sigma^3 ) ~~,~~  \S^{2 \,3} ~=~   \fracm 12 \,(\sigma^2 
\otimes \sigma^1 )    ~~,~~   \S^{3 \,1} ~=~    \fracm 12 \,({\bf I}_2 
\otimes \sigma^2 )   ~~,~~ 
 } \label{Spn-matrX}
\ee
and it is easily seen that these form an SU(2) algebra.  In our discussions, 
we often refer to this as the SU${}_{\a}$(2) algebra.   

However, using the gamma matrices described above, it is possible to 
construct a second SU(2) algebra that we denote as the SU${}_{
\b}$(2) algebra.  The generators of SU${}_{\b}$(2) algebra take the explicit 
forms
\be {
\fracm 12 \,( \sigma^3 \otimes \sigma^2  ) =-  i \, \fracm 12 \,\gamma^0
 ~~,~~ \fracm 12 \,(\sigma^1 \otimes \sigma^2 ) = - \, \fracm 12 \,\gamma^5
~~,~~ \fracm 12 \, (\sigma^2 \otimes {\bf I}_2 )  =  \fracm 12 \, \gamma^0 
\, \gamma^5  ~~. 
 } \label{SU2b-matrX}
\ee
Due to the defining properties of the gamma matrices, these two SU(2)
algebras commute.  This implies that the complete set of sixteen elements
in the covering algebra of the gamma matrices carry a representation of
SU${}_{\a}$(2) $\otimes$ SU${}_{\b}$(2). 

On the other hand, as noted in \cite{G-1} among other places, a four-color 
adinkra in some representation $\cal R$ has a set of L-matrices and R-matrices 
that satisfy the ``Garden Algebra'' conditions
\be  {
\eqalign{
&(\,{\rm R}^{(\cal R)}_\rI\,)_\hi{}^j\>(\, {\rm L}^{(\cal R)}_\rJ\,)_j{}^\hk + (\,{\rm 
R}^{(\cal R)}_\rJ\,)_\hi{}^j\>(\,{\rm L}^{(\cal R)}_\rI\,)_j{}^\hk
= 2 \,  \delta{}_{\rI \, \rJ}  \left(\mathbf{I}_{\rm d}\right)_\hi{}^{\hk}  ~~,~~ \cr
&{\rm L}^{(\cal R)}=~ \left[ \, {\rm R}^{(\cal R)} \, \right]^t=~ \left[ \, {\rm R}^{(\cal R)} \, \right]^{-1}
  ~~.
} }   \label{GrdNAlg}
\ee
From these, one can construct a set of objects called ``adjacency matrices'' 
using a standard definition from graph theory that we can denote by
${\Hat \g}_\rI  $ where
\be {
{\Hat \g}_\rI  ~=~  \left[\begin{array}{cc}
~0 & ~~  \,{\rm L}_\rI\,  \\
~\,{\rm R}_\rI\, & ~0 \\
\end{array}\right]  ~~~.}
 \label{AdjcyMtrX}
\ee
Due to the defining property of the L-matrices and R-matrices
it follows that the quantities ${\Hat \g}_\rI  $ defined this way form
the generators of a four-dimensional Euclidean-signatured Clifford
algebra.  Thus, the quantities defined by ${\Hat \S}_{\rI \, \rJ} \, =\, i/4 
[ \, {\Hat  \g}_{\rI}  \, ~,~ {\Hat \g}_{\rJ}  \, ]$ correspond to a set of
hermitian generators of SO(4).

Owing to the fact that SO(4) is locally isomorphic to SU${}_{\a}$(2) $\otimes$ SU${}_{\b}
$(2), the results of (\ref{G-matrX}) - (\ref{AdjcyMtrX}) imply 
for every adinkra in any representation $\cal R$ with four colors, the 
$N$ $=$ 4 version of the ``Adinkra/$\bm \gamma$-matrix Holography 
Equation'' \cite{KIAS}
\be  {
\eqalign{
({\rm R}^{(\cal R)}_\rI)_\hi{}^j\>({\rm L}^{(\cal R)}_\rJ)_j{}^\hk - (
{\rm R}^{(\cal R)}_\rJ)_\hi{}^j\>({\rm L}^{(\cal R)}_\rI)_j{}^\hk
  &= 2\Big[ \ell^{({\cal R})1}_{\rI\rJ} (\g^2 \g^3){}_{\hi}{}^\hk
   ~+~  \ell^{({\cal R})2}_{\rI\rJ} (\g^3 \g^1){}_{\hi}{}^\hk
   \cr
  &~~~~~~~+~  \ell^{({\cal R})3}_{\rI\rJ} (\g^1 \g^2){}_{\hi}{}^\hk
  ~+~  i\,  {{\Hat \ell}^{(\cal R)}}_{
 \rI\rJ}{}^{1} (\g^0 ){}_{\hi}{}^\hk 
  \cr
 &~~~~~~~+~
  i \, {{\Hat \ell}^{(\cal R)}}_{\rI\rJ}{}^{2} 
 (\g^5){}_{\hi}{}^\hk  ~+~  {{\Hat \ell}^{(\cal R)}}_{\rI\rJ}{}^{3} 
 (\g^0 \g^5){}_{\hi}{}^\hk 
   \Big] 
} }   \label{AdnkG-Mtrx}
\ee
must be valid for some set of constant coefficients $\ell^{({\cal R})I}_{\rI\rJ}$ and 
${\Hat \ell}^{({\cal R})I}_{\rI\rJ}$.   Given that the Adinkra/$\bm \gamma$-matrix 
Holography Equation, for every four-color adinkra, generates a set of matrices 
($\g^0 \g^5$, $\S^{i \, j}$) solely from the data in the adinkra, we simply note the 
equation
\be   {
\g_i ~=~ \e_{i \, j \, k} \, \g^0 \, \g^5 \, \S^{j \, k} ~~~,
}   \label{S8cGMtrX}
\ee
yields the remaining spatial gamma matrices.  After these are constructed we 
multiply by $\g^0$, and then $\g^5$ respectively.  Finally $ \mathbf{I}_{4} $ must 
be included the to construct all sixteen elements of the covering algebra.

\subsection{The Generalized Garden Algebra}

~~~~ We can generalize the original notion of a Garden algebra seen in 
\cite{GRana} to the following form. Denoted by $\mathcal{GR}({ d},\, 
N,\, {\ell}_{\m}) $ this generalized algebra is defined by
\be  {
\L_{({\rm I}} \R_{{\rm J})} \= \R_{({\rm I}} \L_{{\rm J})} \= 2 
[{\cal E}({\ell}_{\m}) ]_{{\rm IJ}} \mathbf{I}_{ d} ~~~ 
\text{for}~ {\rm I},\, {\rm J} \= 1,\, 2,\, \ldots,\, N ~,
} \label{ModGRalg}
\ee
where ${\ell}_{\m}$ is the ``direction" of the reduction,  a four-vector
parameter of the algebra on equal footing with $d$ and $N$, and the ${\cal 
E}({\ell}_{\m}) $ determine the `metric' on the space of
SUSY parameters.

In the following, we shall show that \textit{reduction along any spacetime axis satisfies 
the generalized Garden algebra}. Finally, some explicit examples of the values of 
this SUSY parameter space matric for $\mathcal{GR}(4,\, 4,\, \ell_{\m})$ are
\be \eqalign{
[{\cal E}({\cal T}_{\mu}) ]_{\rm IJ}  ~=  ~ [ \mathbf{I}_{4}  ]_{\rm IJ}
 ~~~~~~~\,\,~&,~~~~
[{\cal E}({\cal X}_{\mu}) ]_{\rm IJ}  ~= ~  [\sigma^3 \otimes \sigma^3  ]_{\rm IJ}
~~, \cr
[{\cal E}({\cal Y}_{\m}) ]_{\rm IJ} ~=~  [\sigma^1 \otimes \mathbf{I}_{2}  ]_{\rm 
IJ} ~~&,~~~~
[{\cal E}({\cal Z}_{\mu}) ]_{\rm IJ}  ~=~ - \, [\sigma^3 \otimes \sigma^1  ]_{\rm IJ}  
~~, } 
{~}   \label{ModGRalg1}
\ee
with the eigenvalues of $\ce(\ct_\mu)$ all being +1 and the eigenvalues of $\ce(\cx_\mu)$, $\ce(\cy_\mu)$, and $\ce(\cz_\mu)$ being $\pm 1$ (each doubly degenerate).  We also see the trace of ${\cal 
E}({\cal T}_{\mu})$ is equal to four, and the traces of ${\cal E}({\cal X}_{\mu})$, 
${\cal E}({\cal Y}_{\mu})$, and ${\cal E}({\cal Z}_{\mu})$ are all equal to zero.

For the case of ${\cal E}({\cal X}_{\mu})$, this is precisely an example of the 
structure presented by Kuznetsova, Rojas, and Toppan \cite{EtAlToppan}
who advocated considering this type of metric in the parameter space of the
1D, $N$-extended SUSY QM systems.  However, in the case of ${\cal E}({
\cal Y}_{\mu})$, and ${\cal E}({\cal Z}_{\mu})$, we see the metric in the
SUSY parameter space can be even more general than they proposed.
For general values of the angles $\a$, $\b$, and $\g$, one finds:
\be \eqalign{
[{\cal E}({\ell}_{\mu}) ]_{\rm IJ}  \= \
&\cos \a \, [{\cal E}({\cal T}_{\mu}) ]_{\rm IJ}  \ + \  \sin \a \, \sin \b \, \cos \g \,
[{\cal E}({\cal X}_{\mu}) ]_{\rm IJ}  \ + \   \cr
& \sin \a \, \sin \b \, \sin \g \, [{\cal E}({\cal Y}_{\mu}) ]_{\rm IJ}  \ + \ 
\sin \a \, \cos \b \, [{\cal E}({\cal Z}_{\mu}) ]_{\rm IJ}   ~~~.
}   \label{ModGRalg2}
\ee
However, we also find a surprising result.  If we impose the condition
\be {
	(\L_{\rm I})^t \= [{\cal E}({\ell}_{\m}) ]_{\rm IJ} (\R_{\rm J}) ~~~ ,
} \label{DuLty}
\ee
as the natural generalization from the temporal 0-brane reduction, this forces
the angle $\a$ to either take the value of  $\a$ $=$ 0, $\p/2$, or  $\a$ $=$ $\p$!   

Let us introduce one other notation in order to illustrate another result.
We can write
\be
{\cal E}(n^1) \= {\cal E}({\cal X}_{\mu}) ~~,~~ {\cal E}(n^2) \= {\cal E}({\cal Y
}_{\mu}) ~~,~~ {\cal E}(n^3) \= {\cal E}({\cal Z}_{\mu}) ~~.
\label{E1s}
\ee
It now follows from (\ref{Spn-matrX}) that we have the nice algebraic relations
\be
\left[ {\cal E}(n^i),\, {\cal E}(n^j)\right]_{\rm IJ} \=  i\, 4 \, (\Sigma^{i \, j})_{\rm IJ}
~~~.
\label{Es}
\ee

Let us note that we can define
\be
 V_{\rm IJ} ~=~ \fracm 12 \left[  \, \L_{\rm I} \R_{\rm J} 
~-~ \L_{\rm I} \R_{\rm J} \, \right]  ~~,~~
 {\Tilde V}_{\rm IJ} ~=~ \fracm 12 \left[ \,  \R_{\rm I} \L_{\rm J} 
~-~ \R_{\rm I} \L_{\rm J} \, \right]   ~~~.
\label{Vmtrx}
\ee
and from these it follows that
\be  \eqalign{
\left[   V_{\rm IJ} ~,~ V_{\rm KL}  \right] ~&=~  [{\cal E}({\ell}_{\mu}) ]_{\rm IK} 
 V_{\rm JL}  -
[{\cal E}({\ell}_{\mu}) ]_{\rm JK}  V_{\rm IL} - 
[{\cal E}({\ell}_{\mu}) ]_{\rm IL}  V_{\rm JK}   + 
[{\cal E}({\ell}_{\mu}) ]_{\rm JL}  V_{\rm IK}  ~~~,   \cr
\left[   {\Tilde V}_{\rm IJ} ~,~ {\Tilde V}_{\rm KL}  \right] ~&=~ 
[{\cal E}({\ell}_{\mu}) ]_{\rm IK} 
 {\Tilde V}_{\rm JL}  - 
[{\cal E}({\ell}_{\mu}) ]_{\rm JK}  {\Tilde V}_{\rm IL} - 
[{\cal E}({\ell}_{\mu}) ]_{\rm IL}  {\Tilde V}_{\rm JK}  + 
[{\cal E}({\ell}_{\mu}) ]_{\rm JL}  {\Tilde V}_{\rm IK}  ~~~.
} \label{Vmtrx2}
\ee
where $\ell$ is picked along each of the coordinate axes.  These commutator
algebras clearly do not describe SO(4) for $\a$ $\ne$ 0.  This is turn implies 
that the adjacency matrices also do {\em {not}} form a representation of SO(4)
for $\a$ $\ne$ 0.

In closing, we note that each supermultiplet when subjected to 0-brane reduction
is associated with its own adjacency matrices.  For distinct multiplets, these
adjacency matrices are different.  In the following we will prove that the results 
in (\ref{DuLty}) and (\ref{Es}) hold independent of which multiplet 
on which the evaluation is made.


$~~~~$ \section{Reduction of the Chiral Multiplet}
\label{section3}

$~~~~$ In order to investigate in concrete detail the procedure of general 0-brane 
reduction and any subtleties that arise along the way, the standard 4D, $\cn$ = 1 
chiral multiplet is considered. Using the conventions of \cite{G-1}, the Lagrangian 
that is invariant with respect to supersymmetry variations takes the form
\be {
\eqalign{
 \mathcal{L} = &  -\frac{1}{2} \partial_{\mu}A \partial^\mu A ~-~ \frac{1}{2} \partial_{
 \mu}B \partial^\mu B  ~+~ \frac{1}{2} i (\gamma^\mu)^{bc} {\psi}_b \partial_\mu {\psi
 }_c ~+~ \frac{1}{2} F^2 + \frac{1}{2} G^2
} } \label{CM1}
\ee
with the SUSY transformation laws in component form  realized by the supercovariant 
derivative ${\rm D}_a$ operators acting on the fields as
\be  \eqalign{
	{\rm D}_a A & \= \psi_a ~~~~~~~~~~~~~~~,~~~~
	{\rm D}_a B  \= i\, ( \gamma^5 )_a{}^b \psi_b ~~~~~~~~~~~~~~,\cr
	{\rm D}_a \psi_b & \= i\, ( \gamma^\mu )_{ab} \,\partial_\mu A - ( \gamma^5 
	\gamma^\mu )_{ab} \,\partial_\mu B - i C_{ab} F + ( \gamma^5 )_{ab} G \,~, \cr
	{\rm D}_a F & \= ( \gamma^\mu )_a{}^b \,\partial_\mu \psi_b ~~~~, ~~~~
	{\rm D}_a G  \= i\, ( \gamma^5 \gamma^\mu )_a{}^b \,\partial_\mu \psi_b ~~~~~~\,~,
}
\label{CM2}
\ee
in four dimensions.
Under the 0-brane reduction where $\partial_{\mu}$ = $\ell_{\mu} \pa_{\t}$ these become
\be \eqalign{
	{\rm D}_a A & \= \psi_a ~~~~~~~~~~~~~~~,~~~~
	{\rm D}_a B  \= i\, ( \gamma^5 )_a{}^b \psi_b ~~~~~~~~~~~~~~,\cr
	{\rm D}_a \psi_b & \= i\, (\gamma \cdot \ell )_{ab} \,\partial_{\tau} A - ( \gamma^5 
	\gamma \cdot \ell)_{ab} \,\partial_{\tau} B - i C_{ab} F + ( \gamma^5 )_{ab} G \,~, \cr
	{\rm D}_a F & \= ( \gamma \cdot \ell)_a{}^b \,\partial_{\tau} \psi_b ~~~~, ~~~~
	{\rm D}_a G  \= i\, ( \gamma^5 \gamma \cdot \ell )_a{}^b \,\partial_{\tau} \psi_b ~~~~~~\,~,
}
\label{CM3}
\ee
and the Lagrangian becomes
\be {
\eqalign{
 \mathcal{L} = &  -\frac{1}{2} \ell_{\mu} \ell^{\mu} \left[~  \left( \pa_{\t}A \right)^2
 \,+\,   \left( \pa_{\t}B \right)^2~ \right]
   ~+~ \frac{1}{2} i \, (\gamma \cdot \ell)^{bc} \, {\psi}_b \partial_\t {\psi
 }_c ~+~ \frac{1}{2} F^2 + \frac{1}{2} G^2   ~~~.
}  } \label{CM4}
\ee

Furthermore, utilizing the explicit form of the $\g^0$ matrix to find $\eta^{0 \, 
0}$ = $- 1$ and the form of $\ell_{\m}$, this becomes
\be {
\eqalign{
 \mathcal{L} 
 =&~  \frac{1}{2}  \cos(2 \a) \left[~  \left( \pa_{\t}A \right)^2
 \,+\,   \left( \pa_{\t}B \right)^2~ \right]
   ~+~ \frac{1}{2} i \, (\gamma \cdot \ell)^{bc} \, {\psi}_b \partial_\t {\psi
 }_c ~+~ \frac{1}{2} F^2 + \frac{1}{2} G^2 
} \, .
 } \label{CM5}
\ee

A next step involves defining the valise related to these equations.  This
is done by making the re-definitions $F$ $\to$ $\pa_{\t} F$ and $G$ $\to$ 
$\pa_{\t} G$ so that we find
\be \eqalign{
	{\rm D}_a A & \= \psi_a ~~~~~~~~~~~~~~~,~~~~
	{\rm D}_a B  \= i\, ( \gamma^5 )_a{}^b \psi_b ~~~~~~~~~~~~~~,\cr
	{\rm D}_a \psi_b & \= i\, (\gamma \cdot \ell )_{ab} \,\partial_{\tau} A - ( \gamma^5 
	\gamma \cdot \ell)_{ab} \,\partial_{\tau} B - i C_{ab} \partial_{\tau} F + 
	( \gamma^5 )_{ab} \partial_{\tau} G \,~, \cr
	{\rm D}_a F & \= ( \gamma \cdot \ell)_a{}^b \, \psi_b ~~~~, ~~~~
	{\rm D}_a G  \= i\, ( \gamma^5 \gamma \cdot \ell )_a{}^b \, \psi_b ~~~~~~\,~,
}
\label{CM6}
\ee
and the Lagrangian becomes
\be {
\eqalign{
 \mathcal{L} = &  -\frac{1}{2} \ell_{\mu} \ell^{\mu} \left[~  \left( \pa_{\t}A \right)^2
 \,+\,   \left( \pa_{\t}B \right)^2~ \right]
   ~+~i \, \frac{1}{2} (\gamma \cdot \ell)^{bc} \, {\psi}_b \partial_\t {\psi
 }_c \cr
 &{~~~~~~}~+~   \frac{1}{2} \left[~  \left( \pa_{\t}F \right)^2
 \,+\,   \left( \pa_{\t}G \right)^2~ \right]   ~~~.
}  } \label{CM7}
\ee
Implementing the further definitions
\be
 {
\eqalign{
\Phi_i ~=~ \left[\begin{array}{c}
A  \\
B  \\
F  \\
G \\
\end{array}\right]   ~~,~~
 \Psi_{\hat k} ~=~- i \,  \left[\begin{array}{c}
\psi_{1} \\
\psi_{2} \\
\psi_{3} \\
\psi_{4} \\
\end{array}\right]   ~~,
}  } \label{CM8}
\ee
 makes it clear that (\ref{CM6}) can be written in the concise forms
\be
\D_{\rm I} \Phi_i \= i\, (\L_{\rm I})_{i \hat{k}} \Psi_{\hat{k}} \qquad \text{and} 
\qquad \D_{\rm I} \Psi_{\hat{k}} \= (\R_{\rm I})_{\hat{k} i} \pa_{ \tau} \Phi_i 
~~~,
 \label{CM9}
\ee
for some matrices $(\L_{\rm I})$ and $ (\R_{\rm I})$. The Lagrangian then
can be written as:
\be   {
\eqalign{
\mathcal{L} ~=~ &   \frac{1}{2}\, \delta^{i \, j} \, \left( \pa_{\t}\Phi_i \right)
\left( \pa_{\t}\Phi_j \right)  \,- i\, \frac{1}{2} \,  \delta^{{\hat k} \, {\hat l}} 
\, \Psi_{\hat{k}} \,  \partial_\t \Psi_{\hat{l}}  ~~~,
}  } \label{CM10}
\ee
where numerically 
we have
\be
 \delta^{{\hat k} \, {\hat l}} ~=~ - \, (\gamma \cdot \ell)^{{\hat k} \, {\hat l}} 
  \label{CM11}
\ee
for purely time-like reduction $ \ell_{\mu} \ell^{\mu}$ = $- \,1$.  The
condition in (\ref{CM11}) is characteristic of Majorana representations
and we have assumed its use in all our previous discussions of adinkras.

In this section, we have given a detailed discussion of the
steps required to obtain the valise formulation of the chiral
supermultiplet.   Similar discussions could be undertaken
for the vector multiplet starting from its equations
\be \eqalign{
\D_a A_\mu & \,=\, (\gamma_\mu)_a{}^b \lambda_b  ~~~, 
\cr
\D_a \lambda_b & \, =\, - \tfrac{i}{4} [\gamma^\mu,\, 
\gamma^\nu]_{ab} F_{\mu \nu} + (\gamma^5)_{ab} \,{\rm d} 
~~~, \cr
D_a {\rm d} & \, =\,  i (\gamma^5 \gamma^\mu)_a{}^b 
\partial_\mu \lambda_b ~,}    \label{VM}
\ee
or the tensor multiplet starting from its equations
\be \eqalign{
\D_a \varphi & \= \chi_a   ~~~, \cr
\D_a B_{\mu \nu} & \= - \tfrac{1}{4} [\gamma_\mu,\, 
\gamma_\nu]_a{}^b \chi_b  ~~~, \cr
\D_a \chi_b & \= i (\gamma^\mu)_{ab} \partial_\mu 
\varphi - (\gamma^5 \gamma^\mu)_{ab} \epsilon_\mu{
}^{\rho \sigma \tau} \partial_\rho B_{\sigma \tau}  ~~~.
}    \label{TM}
\ee
However, the end point of such discussions is precisely
the same as in (\ref{CM9}) and  (\ref{CM10}) but with 
different L-matrices and R-matrices describing the
distinct supermultiplets.

\section{Coordinate Axis Reductions}

~~~ Explicit reduction of the chiral, vector and tensor supermultiplet in 
a purely time-like direction has been presented before \cite{G-1}.  Once 
the valise form of each supermultiplet is obtained, the supersymmetry 
variations are completely described by the results in (\ref{CM9}).  Thus, 
all that remains is to give the explicit form of the L-matrices and the 
R-matrices. Below these are written in our compact matrix notation as explained in the appendix. A straightforward calculation with the matrices below reveals that our assertion in section \ref{section2} is correct: that Eqs.~(\ref{DuLty}) and~(\ref{Es}) hold independent of multiplet or reduction coordinate. \newline
$~$ \newline
\noindent {\bf {Chiral Multiplet}} \newline
\noindent
Reduction for $\ell$ = $\cal T$ gives the following L and R matrices
\be  \eqalign{ {~~~~}
\L_1 \,&=\, (1\, {\bar 4}\, 2\, {\bar 3}) ~~,~~ \L_2  \,=\, (2\, 3\, {\bar 1}\, {\bar 4}) 
~~,~~\,  \L_3   \,=\, (3\, {\bar 2}\, {\bar 4}\, 1)~~,~~   \L_4  \,=\, (4\, 1\, 3\, 2)  ~~,
\cr
\R_1 \,&=\, (1 \, 3 \, {\bar 4} \,{\bar 2} ) ~~,~~ \R_2  \,=\, ({\bar 3} \,1 \, 2 \, {\bar 4}) 
~~,~~  \R_3   \,=\, (4\, {\bar 2} \,1\,{\bar 3})~~,~~   \R_4  \,=\, (2\, 4\, 3\, 1)  ~~.
} \label{LRmtrx0c}
\ee
Reduction for $\ell$ = $\cal X$ gives the following L and R matrices
\be  \eqalign{ {~~~~}
\L_1 \,&=\, (1 \,{\bar 4} \,2 \,{\bar 3} ) ~~,~~ \L_2  \,=\, (2 \,3 \, 1 \, 4 ) 
~~,~~\,  \L_3   \,=\, (3\, {\bar 2} \,4 \,{\bar 1} )~~,~~   \L_4  \,=\, (4\, 1\, 3\, 2)  ~~,
\cr
\R_1 \,&=\, (1 \, 3 \, {\bar 4} \,{\bar 2} ) ~~,~~ \R_2  \,=\, ({\bar 3} \,{\bar 1} \, {\bar 2} \, {\bar 4} ) 
~~,~~  \R_3   \,=\, (4\, 2\, {\bar 1}\,{\bar 3} )~~,~~   \R_4  \,=\, (2\, 4\, 3\, 1)  ~~.
} \label{LRmtrx1c}
\ee
Reduction for $\ell$ = $\cal Y$ gives the following L and R matrices
\be  \eqalign{ {~~~~}
\L_1 \,&=\, (1 \,{\bar 4} \,{\bar 4} \, 1 ) ~~,~~ \L_2  \,=\, (2 \,3 \, 3 \, 2 ) 
~~,~~\,  \L_3   \,=\, (3\, {\bar 2} \, 2 \, {\bar 3} ) ~~~,~~   \L_4  \,=\, (4\, 1\, {\bar 1}\, 
{\bar 4}  )  ~~,  \cr
\R_1 \,&=\, (3 \, {\bar 2} \, 2 \,{\bar 3} )^t \, ,~~ \R_2  \,=\, (4 \, 1 \, {\bar 1} \, {\bar 4} )^t 
\,,~~  \R_3   \,=\, (1\, {\bar 4} \,{\bar 4}\, 1 )^t \,,~~   \R_4  \,=\, (2\, 3\, 3\, 2 )^t  ~.
} \label{LRmtrx2c}
\ee
Reduction for $\ell$ = $\cal Z$ gives the following L and R matrices
\be  \eqalign{ {~~~~}
\L_1 \, &=\, (1 \,{\bar 4} \, 1 \, 4 ) ~~,~~ \L_2  \,=\, (2 \,3 \, {\bar 2} \, 3 ) 
~~,~~\,  \L_3   \,=\, (3\, {\bar 2} \, 3 \, 2 ) ~~~,~~   \L_4  \,=\, (4\, 1\, {\bar 4} \, 1 )  ~~,  \cr
\R_1 \, &=\, ({\bar 2} \, {\bar 3} \, 2 \,{\bar 3} )^t \, ,~~ \R_2  \,=\, ({\bar 1} \, 4 \, {\bar 1} \, {\bar 4} )^t 
\,,~~  \R_3   \,=\, (4\, 1\, {\bar 4} \, 1 )^t ~,~~   \R_4  \,=\, (3\, {\bar 2}\, 3\, 2 )^t  ~.
} \label{LRmtrx3c}
\ee
\newline
$~$ \newline
\noindent {\bf {Vector Multiplet}} \newline
\noindent
Reduction for $\ell$ = $\cal T$ gives the following L and R matrices
\be   \eqalign{ {~~~~}
\L_1 \,&=\, (2 \,{\bar 4} \,1 \,{\bar 3} ) ~~,~~ \L_2  \,=\, (1 \,3 \,{\bar 2} \, {\bar 4} ) 
~~,~~\,  \L_3   \,=\, (4\, 2 \, 3 \,1 )~~,~~   \L_4  \,=\, (3\, {\bar 1}\, {\bar 4}\, 2  )  ~~,
\cr
\R_1 \,&=\, (3 \, 1 \, {\bar 4} \,{\bar 2} ) ~~,~~ \R_2  \,=\, (1 \, {\bar 3} \, 2 \, {\bar 4} ) 
~~,~~  \R_3   \,=\, (4\, 2 \,3\, 1 )~~,~~   \R_4  \,=\, ({\bar 2}\, 4\, 1\, {\bar 3} )  ~~.
} \label{LRmtrx0v}
\ee 
Reduction for $\ell$ = $\cal X$ gives the following L and R matrices
\be \eqalign{ {~~~~}
\L_1 \, &= ( \bar{2} \, \bar{4} \, 1 \bar{3} ) \,~~,~~\,  \L_2 \, =\, (1 \, 3 \, \bar{2} \, 4) 
~~,~~\, \L_3 \, =\, (4 \, 2 \, 3\,  \bar{1}  ) ~~,~~ \L_4 \, =\, (\bar{3} \,  \bar{1}\,  \bar{4} \,
2 ) ~~,  \cr
\R_1 \, &=\, (3\, \bar{1}\,  \bar{4}\, \bar{2} ) ~~,~~  \R_2 \, =\, ( \bar{1} \, 3 \, \bar{2} \, 
\bar{4} ) ~~,~~  \R_3 \, =\,  ( 4 \, \bar{2}\,  \bar{3} \,  \bar{1} )  ~~,~~ \R_4 \, =\, 
( \bar{2} \, 4  \, \bar{1} \, \bar{3} ) ~~.
} \label{LRmtrx1v}
\ee
Reduction for $\ell$ = $\cal Y$ gives the following L and R matrices
\be \eqalign{ {~~~~}
\L_1 \, &= \,(\bar{2} \, 2 \, 1\, 1) ~~~,~~  \L_2 \, = \,(1 \,1\, \bar{2} \, 2) ~~\,,~~  \L_3 \, = \,(4 \,4 \,3 \,
\bar{3}) ~~~,~~  \L_4 \, = \,(\bar{3}\, 3\,  \bar{4}\, \bar{4}) ~~~,  \cr
\R_1 \, &= \,(4 \,4 \, 3\, \bar{3})^t ~,~~  \R_2 \, = \,(\bar{3} \,3\, \bar{4} \, \bar{4})^t ~,~~  
\R_3 \, = \,(\bar{2} \, 2 \, 1 \, 1)^t ~,~~ \R_4 \, = \,(1 \,1 \,\bar{2} \, 2)^t ~.
} \label{LRmtrx2v}
\ee
Reduction for $\ell$ = $\cal Z$ gives the following L and R matrices
\be \eqalign{ {~~~~}
\L_1 \, &=\, (\bar{2} \,2 \,\bar{4} \, 4) ~\,~,~~  
\L_2 \, = \,(1\, 1 \, 3 \, 3) ~~~,~~  \L_3 \, = \,(4 \, 4 \, 2 \, 2) 
~~~,~~ \L_4 \, = \,(\bar{3} \, 3 \,\bar{1} \, 1) ~~~,  \cr
\R_1 \, &=\,  ({\bar 1} \,{\bar 1} \, {\bar 3} \, {\bar 3})^t ~,~~ 
 \R_2 \, = \, (2 \, {\bar 2} \, 4\, {\bar 4})^t ~,~~  \R_3 \, = \,
(\bar{3} \, 3 \, \bar{1} \, 1)^t ~,~~ \R_4 \, = \,(4 \,4 \,2 \,2)^t ~.
} \label{LRmtrx3v}
\ee
\newline
\noindent {\bf {Tensor Multiplet}} \newline
\noindent
Reduction for $\ell$ = $\cal T$ gives the following L and R matrices
\be   \eqalign{ {~~~~}
\L_1 \,&=\, (1 \,{\bar 3} \,{\bar 4} \,{\bar 2} ) ~~,~~ \L_2  \,=\, (2 \,4 \,{\bar 3} \, 1 ) 
~~,~~\,  \L_3   \,=\, (3\, 1 \, 2 \,{\bar 4} )~~,~~   \L_4  \,=\, (4\, {\bar 2}\, 1\, 3  )  ~\,~,
\cr
\R_1 \,&=\, (1 \, {\bar 4} \, {\bar 2} \,{\bar 3} ) ~~,~~ \R_2  \,=\, (4 \, 1\, {\bar 3} \, 2 ) 
~~,~~  \R_3   \,=\, (2 \,3\, 1 \,{\bar 4} )~~,~~   \R_4  \,=\, (3 \, {\bar 2}\, 4\, 1 )  ~~.
} \label{LRmtrx0t}
\ee 
Reduction for $\ell$ = $\cal X$ gives the following L and R matrices
\be \eqalign{ {~~~~}
~\L_1  \, &= \, (1 \, 3 \, \bar{4} \, 2) ~~,~~  \L_2  \, = \, (2 \, 4\, \bar{3}\, 1) ~~,~~\,   \L_3  \, = \, 
(3 \, 1 \, 2 \, \bar{4}) ~~,~~   \L_4  \, = \, (4 \, 2 \, 1 \, \bar{3}) ~\,~,  \cr
~\R_1  \, &= \, (1 \, 4 \, 2 \, \bar{3}) ~~,~~   \R_2  \, = \, (\bar{4} \, \bar{1} \, 3\,  \bar{2}) ~~, ~~
  \R_3  \, = \, (\bar{2} \,\bar{3} \, \bar{1} \, 4) ~~,~~   \R_4  \, = \, (3 \, 2 \, \bar{4} \, 1)
  ~~.
} \label{LRmtrx1t}
\ee 
Reduction for $\ell$ = $\cal Y$ gives the following L and R matrices
\be \eqalign{ {~~~~}
\L_1 \, &=\, (1 \, 1 \, 2 \, 2) ~~\,,~~   \L_2  \, = \, (2 \, \bar{2} \,  \bar{1} \, 1) ~~~,~~ \L_3  \, = \, 
(3 \, \bar{3} \, 4\,  \bar{4}) ~~~,~~   \L_4  \, = \, (4 \, 4 \, \bar{3}  \,\bar{3}) ~~~,   \cr
\R_1 \, &= \, (3\, \bar{3} \, 4  \, \bar{4})^t ~,~~   \R_2  \, = \, (4 \, 4 \,\bar{3}\, \bar{3})^t ~,~~
\R_3  \, = \, (1 \, 1 \, 2 \, 2)^t ~,~~   \R_4  \, = \, (2 \, \bar{2} \,  \bar{1} \, 1)^t 
 ~.
} \label{LRmtrx2t}
\ee 
Reduction for $\ell$ = $\cal Z$ gives the following L and R matrices
\be \eqalign{ {~~~~}
\L_1 \, &=\, (1 \,1 \, \bar{3} \, \bar{3}) ~\,~,~\,~   \L_2  \, = \, (2\,  \bar{2} \, 4 \, \bar{4}) ~~~,~~  \L_3  \, = \, (3\,
 \bar{3}\, 1 \, \bar{1}) ~~\,,\,~~  \L_4  \, = \, (4 \, 4 \, \bar{2}\, \bar{2}) ~\,~,  \cr
\R_1 \, &= \, ({\bar 2} \, 2 \, {\bar 4}\, 4)^t ~,~~   \R_2  \, = \,  ({\bar 1} \,{\bar 1}\, 3 \, 3)^t ~,   ~~
\R_3  \, = \, (4 \,4\, \bar{2} \, \bar{2})^t ~,~~  \R_4  \, = \, (3\,  \bar{3} \,1\, \bar{1})^t ~.
} \label{LRmtrx3t}
\ee 


\section{Complications of Light-like 0-Brane Projection}

$~~~~$ 
There is one other subtlety that we need to address shortly.  We start by noting that a simple calculation using the parametrization in terms of
$\a$, $\b$, and $\g$ shows
\be
\ell \, \cdot \, \ell ~=~\ell_{\m} \, \eta^{\m \, \n} \, \ell_{\n} ~=~ - \, \cos(2 \a)
~~~.
 \label{InnRProd}
\ee
Clearly the constant four-vector $\ell_{\m}$ defines three distinct regimes 
for 0-brane reduction.
\be
-\, \cos(2 \a) ~\propto~  \left\{
\begin{array}{c}
~< \, 0 ~,~~{\text{for~time-like}} ~ \ell_{\m} \\
~= \, 0  ~,~~{\text{for~light-like}} ~ \ell_{\m}\\
~\,> \, 0  ~,~~{\text{for~space-like}} ~ \ell_{\m}\\
\end{array}\right\}
\ee
It turns out that light-like reduction has subtle differences from the other
two cases.  We will briefly discuss this here.

Given any fixed four-vector of the form that appears in (\ref{DerV-t}), it is 
always possible to construct a second such four-vector from this one.  The 
second four-vector is linearly independent of the first and obtained from it 
by reversing the signs of all the spatial components of the first such four-vector.  
For light-like 0-brane reduction this is important for reasons having to do 
with dynamics.

Thus, even for a four-vector, denoted by $\ell_{\m}^{(\pp)}$, that satisfies 
$\ell^{(\pp)} \cdot \ell^{(\pp)} \,=$ 0, there exists a second four-vector, denoted
by $\ell_{\m}^{(\mm)}$, constructed as described immediately above.
These two satisfy the conditions
\be {
\ell_{\m}^{(\pp)} \, \eta^{\m \, \n} \, \ell_{\n}^{(\pp)} ~=~ 0 ~,~
\ell_{\m}^{(\mm)} \, \eta^{\m \, \n} \, \ell_{\n}^{(\mm)} ~=~ 0 ~,~
\ell_{\m}^{(\pp)} \, \eta^{\m \, \n} \, \ell_{\n}^{(\mm)} ~=~ -1  ~~~,
} \label{LghtVecs}
\ee
using the parametrization provided by (\ref{DerV-t}).

Under light-like 0-brane reduction we use $\pa_{\m}$ = $ \ell_{\mu
}^{(\pp)}\, \pa_{\pp} $ + $ \ell_{\mu}^{(\mm)}\, \pa_{\mm} $ where 
$ \pa_{\pp} $ and $ \pa_{\mm} $ are independent derivative
operators.  Since they are independent, this is no longer a 0-brane
reduction because one is actually reducing to a 1-brane where the 
two vector fields $ \pa_{\pp} $ and $ \pa_{\mm} $ (in the mathematical 
sense) generate motion on a world-sheet.  In this circumstance the 
algebra of SUSY generators take the form
\be {
\eqalign{
\{ \, {\rm Q}_a  \,,\,  {\rm Q}_b  \,  \} \,  
~&=~  i\, 2 \,  \, (\gamma \cdot \ell^{\pp}){}_{a \,b} \, \pa_{\pp} \,+\,   
 i\, 2 \,  \, (\gamma \cdot \ell^{\mm}){}_{a \,b} \, \pa_{\mm}
~~~.
}   }  \label{SUSYalg2}
\ee

The form of the SUSY variations in (\ref{CM6}) do not exhibit any pathological 
behavior for any values of the parameters $\a$, $\b$, and $\g$.   The Lagrangian 
is a very different story in this regard and picks out a special value in the parameter 
space that requires a more careful analysis.  

As long as $ \ell_{\mu} \ell^{\mu}$ $\ne $ 0 (or alternately $\a$ $\ne$ $\p$/4
or $\a$ $\ne$ $3\p$/4), the Lagrangian in (\ref{CM5}) shows that by re-scaling 
the bosons and possibly reversing the sign of the $\t$-derivative all such 
reduced action can be brought to the same form.  However, if $ \ell_{\mu} 
\ell^{\mu}$ $=$ 0, then the form of (\ref{CM5}) shows there is a complication --- the 
$A$ and $B$ terms are absent from the Lagrangian and the off-shell equality 
in the number of bosonic versus fermionic degrees of freedom is lost.

In the case of a light-like $\ell$-parameter, the reduction must be done
to a 1-brane (not a 0-brane) or equivalently to a world-sheet.   But it
should be clear that it is the requirement of being able to write an appropriate
Lagrangian, \textit{i.e.} the dynamics, that forces these changes in analysis.

For these we note $\pa_{\m}$ = $ \ell_{\mu
}^{(\pp)}\, \pa_{\pp} $ + $ \ell_{\mu}^{(\mm)}\, \pa_{\mm} $ so the Lagrangian 
we then find is
\be {
\eqalign{
 \mathcal{L} = &~  \left(  \partial_{\pp}A  \right)  \left(  \partial_{\mm}A  \right)  ~+~ 
   \left(  \partial_{\pp}B  \right)  \left(  \partial_{\mm}B  \right)    \cr
 &~+~ \frac{1}{2} i (\gamma \cdot \ell^{(\pp)} )^{bc} {\psi}_b \partial_\pp {\psi
 }_c ~+~ \frac{1}{2} i (\gamma \cdot \ell^{(\mm)} )^{bc} {\psi}_b \partial_\mm {\psi
 }_c    \cr
 &~+~ \frac{1}{2} F^2 + \frac{1}{2} G^2
}  } \label{CM12}
\ee
and the SUSY
transformation laws in component form realized via the supercovariant 
derivative ${\rm D}_a$ operator acting on the fields now become
\begin{align*}
	{\rm D}_a A & \= \psi_a ~~~~~~~~~~~~~~~,~~~~
	{\rm D}_a B  \= i\, ( \gamma^5 )_a{}^b \psi_b ~~~~~~~~~~~~~~,\\[4pt]
	{\rm D}_a \psi_b & \= i\, (\gamma \cdot \ell^{(\pp)} )_{ab} \,\partial_\pp A
	+   i\, (\gamma \cdot \ell^{(\mm)} )_{ab} \,\partial_\mm A \\[4pt] 
	 &{~~~~~}- ( \gamma^5 
	\gamma \cdot \ell^{(\pp)} )_{ab} \,\partial_\pp B  
	- ( \gamma^5 
	\gamma \cdot \ell^{(\mm)} )_{ab} \,\partial_\mm B 
	 \\[4pt] 
	&{~~~~~}- i C_{ab} F + ( \gamma^5 )_{ab} G \,~, \\[4pt]
	{\rm D}_a F & \= ( \gamma \cdot \ell^{(\pp)}  )_a{}^b \,\partial_\pp \psi_b +
	 ( \gamma \cdot \ell^{(\mm)}  )_a{}^b \,\partial_\mm \psi_b
	~~~~,  \\[4pt]
	{\rm D}_a G & \= i\, ( \gamma^5 \gamma \cdot \ell^{(\pp)}  )_a{}^b \,\partial_\pp \psi_b +
	i\, ( \gamma^5 \gamma \cdot \ell^{(\mm)}  )_a{}^b \,\partial_\mm \psi_b
	~~~~~~\,~.
\end{align*}
\be
{~} \label{CM13}
\ee
So one can see the distinctive features that light-like 0-brane reduction possesses
in comparison to the other cases.  In fact, light-like 0-brane reduction is inconsistent
with maintaining the off-shell equality of bosons and fermions.


\section{Conclusions}

$~~~~$ Though we have investigated the 0-brane reduction only of the minimal off-shell
4D, $\cal N$ $=$ 1 supermultiplets, we expect our results to hold more generally.   One of the most
fascinating revelations of this work, is how a purely time-like 0-brane reduction is 
distinguished from all others. In the following, we give a detailed discussion of this point.

The results for the various on-axis reductions of the chiral supermultiplet (\ref{LRmtrx0c} - \ref{LRmtrx3c}),
vector supermultiplet (\ref{LRmtrx0v} - \ref{LRmtrx3v}), and tensor supermultiplet (\ref{LRmtrx0t} - \ref{LRmtrx3t}), can all be substituted into equations of section \ref{section2} to show that these latter
equations are valid for each of the supermultiplets, independent of which set of L-matrices
and R-matrices are used for the various supermultiplets.

 It is useful to reflect on analogous results for the group SU(3). The standard 
Gell-Mann matrices $\l_A$ are known to satisfy a commutator algebra
\be
 [ \,   \l_A  ~,~ \l_B \, ] ~=~ i \, f_{A \, B}{}^C \,  \l_C
\ee
for a very well-known set of structure constants $ f_{A \, B}{}^C$.  Upon taking
the complex conjugate of this equation, one learns that $- (\l_A)^*$ satisfies the
same equation.   The quantities $\fracm 12 \l_A$ are the SU(3) generators 
acting on the quark states while $- \fracm 12 (\l_A)^*$ are the SU(3) generators 
acting on the anti-quark states.  This is similar to fact the the distinct sets of L-matrices
and R-matrices satisfy the same algebra. 

  We know from the results in \cite{KIAS} that for temporal reduction,
  the Adinkra/$\g$-matrix Holographic Equation (\ref{AdnkG-Mtrx}) holds.   The discussion
  surrounding (\ref{Vmtrx}) and (\ref{Vmtrx2}) shows this not the case if
  $\a$ $\ne$ 0.  As 
well, from \cite{permutadnk}, using temporal reductions, we have found 
a possible realization of the Hodge star operator of 4D, $\cal N$ = 1 
supermultiplets which appears to have a `shadow'  realization acting 
on 1D, $N$ = 4 adinkras.  This last point is critical for the definition
of a class structure to exist that allows a purely one dimensional distinction 
between adinkras associated with the chiral, vector, and tensor supermultiplets, 
independent of any reference to higher dimensions.  It is not at all clear 
whether these results can be extended to reductions that possess values 
of the angle $\a$ $\ne$ 0. 

The 0-brane reduction along the temporal direction for supermultiplets 
and adinkras appears to be distinguished as being different from other 
directions.   

 \vspace{.05in}
 \begin{center}
 \parbox{4in}{{\it ``There is no difference between time
 and of the three di- $~~$ mensions of space except that our consciousness
 moves  $~~$ along with it.''}\,\,-\,\, The Time Traveller $~~~~~~~~~~~~~~~~~~~~~~$
 $~~~~~~~~~~~~~~~~~~~~~~~~~~~$
 from H.\ G.\ Wells' \textit{The Time Machine}}  
 \end{center}
 
  \noindent
{\bf Acknowledgements}\\[.1in] \indent
This work was partially supported by the National Science Foundation grants 
PHY-0652983 and PHY-0354401. This research was also supported in part by the 
endowment of the John S.~Toll Professorship and the University of Maryland Center for 
String \& Particle Theory.    We also thank T. H\"ubsch for the alternative description suggested below and F. Toppan for pointing out an incorrect reference in a previous version of the paper.
$$~~$$ $$~~$$

\noindent
{\bf Added Note In Proof}\\[.1in] \indent
T. H\" ubsch has suggested an alternative prescription for 0-brane projection
that when utilized modifies our results.  In the discussion presented in section
\ref{section3}, the 0-brane projection for the superspace covariant derivative takes
the form
\be
{\rm D}_a  ~~  \to~~\D_{\rm I} ~~~.
\ee
The alternate suggestion is to use a projection of the form
\be
{\rm D}_a  ~~  \to~~ \left[  {\mathcal F}(\ell) \right]_{\rm I}{}^{\rm J} \, \D_{\rm J}
~~~,
\ee
where the factor $ \left[  {\mathcal F}(\ell) \right]_{\rm I}{}^{\rm J} $ has the
form of a particular Lorentz transformation.  If $\ell$ is in the forward lightcone,
then this Lorentz transformation may be chosen so as to transform $\ell_{\mu}$
to the purely time-like axis.  As there exists no Lorentz transformation that 
can transform a light-like $\ell_{\mu}$ to a time-like $\ell_\mu$, it is clear that light-like 0-brane 
reduction would remain distinct.  Similarly, there is also no Lorentz transformation 
that can transform a space-like $\ell_{\mu}$ to a time-like $\ell_\mu$, so space-like 
0-brane reduction would remain distinct as well.

Under this alternate prescription for 0-brane reduction, the only effectively
distinct cases correspond to the angular parameter $\alpha$ in (\ref{DerV-t}) being
restricted to the values of $0$, $\pi/4$, $\pi/2$, $3 \pi/4$, or $\pi$.  The
first and last of these are time-like, the second and fourth are light-like,
and the third is space-like.

The result of (\ref{AdnkG-Mtrx}) can only occur for time-like 0-brane reduction.  
That `bridge' between four-color adinkras and the $\gamma$-matrices of 
SO(1,\,3) only exists for time-like 0-brane reductions.  So all time-like 0-brane 
reductions are equivalent to a purely temporal reduction and this case remains 
distinct from the other cases.

\appendix

\section{Defining A Compact Matrix Notation}

$~~~~$ In order to present our result most compactly, we have used a notation that we
initiated in the work of \cite{permutadnk}.  In this work, we showed that since
the L-matrices and R-matrices associated with purely time-light 0-brane reduction
may be thought of as being constructed by multiplying a set of unimodal matrices 
of order two by elements of $S_4$, the permutation group of four elements,
one of the standard notations for permutations may be adapted to our discussions.

This notation is simply a nice way to represent our matrices, most easily defined through 
example:
\begin{equation}
	\left( \begin{array}{cccc}
		0 & -1 & 0 & 0 \\
		0 & 0 & 0 & -1 \\
		1 & 0 & 0 & 0 \\
		0 & 0 & -1 & 0
	\end{array} \right) ~\equiv~ \left(\bar{2}\, {\bar 4}\, 1 \, \bar{3} \right) \,.
\end{equation}
In the expression $(ijkl)$, $i$ represents the column in which the non-zero entry 
of the first row sits; $j$ represents the same but for the second row, and so on. A 
bar over the index signifies that the element in that spot should be $-1$ instead 
of $+1$.  Finally, we use the notation $\left(\bar{2}\, {\bar 4}\, 1 \, \bar{3} \right)^t$
to refer to the transpose of the matrix in (A.1).
$$~~$$

\end{document}